\documentclass[twocolumn,bookmarks=true,pdfnewwindow=true,colorlinks=true,linkcolor=xlinkcolor,citecolor=xlinkcolor,filecolor=xlinkcolor,urlcolor=xlinkcolor,final=true]{aastex631}
\shorttitle{Dust-to-gas ratios in QGs}
\shortauthors{Lorenzon et al.}

\usepackage{graphicx}

\usepackage{siunitx}
\DeclareSIUnit \parsec {pc}
\DeclareSIUnit \year {yr}
\usepackage{amsmath}
\usepackage{bm}
\usepackage{multirow}
\usepackage{booktabs}
\usepackage[utf8]{inputenc}
\usepackage{natbib}
\begin{document}

\title{ALMA reveals diverse dust-to-gas mass ratios and quenching modes in old quiescent galaxies}

\correspondingauthor{G. Lorenzon}
\email{giuliano.lorenzon@ncbj.gov.pl}

\author [0009-0002-8726-8917] {G. Lorenzon}
\affiliation{National Center for Nuclear Research, Pasteura 7, 02-093 Warsaw, Poland}

\author[0000-0001-5341-2162] {D. Donevski}
\affiliation{National Center for Nuclear Research, Pasteura 7, 02-093 Warsaw, Poland}
\affiliation{SISSA, Via Bonomea 265, 34136 Trieste, Italy}

\author[0000-0003-2475-124X] {A. W. S. Man}
\affiliation{Department of Physics \& Astronomy, University of British Columbia, 6224 Agricultural Road, Vancouver, BC V6T 1Z1, Canada}

\author[0000-0002-9948-3916] {M. Romano}
\affiliation{Max-Planck-Institut f\"ur Radioastronomie, Auf dem H\"ugel 69, 53121, Bonn, Germany}
\affiliation{INAF, OAPD, Vicolo dell'Osservatorio, 5, 35122 Padova, Italy}

\author[0000-0001-7160-3632] {K. E. Whitaker}
\affiliation{Department of Astronomy, University of Massachusetts, Amherst, MA 01003, USA}

\author[0000-0002-5615-6018] {S. Belli}
\affiliation{Dipartimento di Fisica e Astronomia, Universit\`a di Bologna, Bologna, Italy}

\author[0000-0001-9773-7479] {D. Liu}
\affiliation{Purple Mountain Observatory, Nanjing, China}

\author[0000-0002-2419-3068] {M. M. Lee}
\affiliation{DTU-Space, Technical University of Denmark, Elektrovej 327, DK-2800 Kgs. Lyngby, Denmark}

\author[0000-0002-7064-4309] {D. Narayanan}
\affiliation{Department of Astronomy, University of Florida, 211 Bryant Space Sciences Center, Gainesville, FL, USA}
\affiliation{Cosmic Dawn Center (DAWN), Copenhagen, Denmark}

\author[0000-0002-7530-8857] {A. Long}
\affiliation{Department of Astronomy, University of Washington, Seattle, WA 98195-1700, USA}
\affiliation{Department of Astronomy, The University of Texas at Austin, 2515 Speedway Boulevard Stop C1400, Austin, TX 78712, USA}

\author[0000-0003-4702-7561] {I. Shivaei}
\affiliation{Centro de Astrobiolog\'ia (CAB), CSIC-INTA, Carretera de Ajalvir km 4, Torrej\'on de Ardoz 28850, Madrid, Spain}

\author[0000-0001-6652-1069] {A. Nanni}
\affiliation{National Center for Nuclear Research, Pasteura 7, 02-093 Warsaw, Poland}
\affiliation{INAF - Osservatorio Astronomico d'Abruzzo, Via Maggini SNC, 64100, Teramo, Italy}

\author[0009-0004-1140-3363] {K. Lisiecki}
\affiliation{National Center for Nuclear Research, Pasteura 7, 02-093 Warsaw, Poland}

\author[0000-0002-0498-8074] {P. Sawant}
\affiliation{National Center for Nuclear Research, Pasteura 7, 02-093 Warsaw, Poland}

\author[0000-0002-9415-2296] {G. Rodighiero}
\affiliation{Dipartimento di Fisica e Astronomia, Universit\`a di Padova, Vicolo dell'Osservatorio, 3, I-35122 Padova, Italy}

\author[0000-0003-4797-5246]{I. Damjanov}
\affiliation{Department of Astronomy and Physics, Saint Mary's University, 923 Robie Street, Halifax, NS, B3H 3C3, Canada}

\author[0000-0002-7016-4532] {Junais}
\affiliation{Instituto de Astrof\'{i}sica de Canarias, V\'{i}a L\'{a}ctea S/N, E-38205 La Laguna, Spain}
\affiliation{Departamento de Astrof\'{i}sica, Universidad de La Laguna, E-38206 La Laguna, Spain}

\author[0000-0003-2842-9434] {R. Dav\'e}
\affiliation{Institute for Astronomy, Royal Observatory, University of Edinburgh, Edinburgh EH9 3HJ, UK}
\affiliation{Department of Physics \& Astronomy, University of the Western Cape, Robert Sobukwe Rd, Bellville, 7535, South Africa}

\author[0000-0003-2606-6019] {C. Pappalardo}
\affiliation{Instituto de Astrof\'isica e Ci\^encias do Espa\c{c}o, Universidade de Lisboa - OAL, Tapada da Ajuda, 1349-018, Lisboa, Portugal}

\author[0000-0001-7964-5933] {C. Lovell}
\affiliation{Institute of Cosmology and Gravitation, University of Portsmouth, Burnaby Road, Portsmouth PO1 3FX, UK}

\author[0000-0001-9626-9642] {M. Hamed}
\affiliation{Centro de Astrobiolog\'ia (CAB), CSIC-INTA, Carretera de Ajalvir km 4, Torrej\'on de Ardoz 28850, Madrid, Spain}

\begin{abstract}
Recent discoveries of dust and molecular gas in quiescent galaxies (QGs) up to $z\sim3$ challenge the long-standing view that the interstellar medium depletes rapidly once star formation ceases, raising key questions of whether dust and gas co-evolve in QGs, and how their depletion links to stellar aging. We present deep Atacama Large Millimeter/submillimeter Array (ALMA) Band~6 continuum and CO(3–2) observations of 17 QGs at $z\sim0.4$ in the COSMOS field. Using the dust-to-molecular gas mass ratio ($\delta_{\rm DGR}$) as a key diagnostic, we trace post-quenching evolution of the cold interstellar medium. Our study triples the number of QGs with direct $\delta_{\rm DGR}$ estimates, constraining 12 systems with stellar population ages of $\sim$5–10 Gyr. For the first time, we show that $\delta_{\rm DGR}$ in QGs ranges from $\sim8\times$ below to $\sim2.5\times$ above the canonical value of $\delta_{\rm DGR}\sim1/100$. Despite uniformly low molecular gas fractions (median $f_{\rm H_2}=M_{\rm H_2}/M_{\star}\sim4.1\%$), QGs follow diverse evolutionary paths: about half exhibit rapid ($\sim700$ Myr) exponential dust decline with age, while the rest show mild decline over $\gtrsim$2 Gyr, maintaining elevated $\delta_{\rm DGR}\gtrsim1/100$. Our results support simulations predictions of dust and molecular gas evolving independently post-quenching, without a preferred quenching mode. This challenges the use of dust continuum as a $\rm H_2$ tracer, implying that quenching cannot be robustly linked to interstellar medium conditions when relying solely on dust or gas.
% OR VERSION (2): ... that the evolutionary fate of QGs can only be assessed by jointly tracing both dust and molecular gas.

%PREVIOUS VERSION BELOW !!: 
%Using key diagnostics as gas-to-stellar, dust-to-stellar and dust-to-gas mass ratios, we study how the cold interstellar medium evolves after quenching. Our study triples the number of QGs with direct dust-to-gas ratio estimates, constraining 10 out of 17 systems with stellar ages of $\sim$5–10 Gyr. For the first time, we reveal dust-to-gas ratios varying by over $2.5\sigma$ both in excess and suppressed compared to the canonical value of $\delta_{\rm DGR} \sim 1/100$. Despite uniformly low molecular gas fractions (median $f_{\rm H_2} \sim 4.1\%$), QGs follow markedly diverse evolutionary paths: roughly half exhibit a rapid ($\sim 700\:\rm Myr$) exponential decline in dust and gas content with stellar age, consistent with passive depletion, while the rest retain substantial dust reservoirs ($\log(M_{\rm dust}/M_{\star}) \gtrsim -3.5$) and elevated dust-to-gas ratios ($\delta_{\rm DGR} \gtrsim 1/100$), declining only mildly over 2 Gyr since quenching.

%Our analysis supports simulation-based predictions that molecular gas and dust may evolve independently after quenching, without a clear preference for a single quenching mode. This challenges the reliability of dust continuum as a gas tracer in QGs and complicates efforts to trace their evolutionary fate without direct measurement of both molecular gas and dust contents.%using only measurements of the molecular gas or dust content.
\end{abstract}


\section{Introduction}
%In the current paradigm of galaxy evolution quiescent galaxies (QGs) have little to no star formation as quenching processes transform their interstellar medium (ISM) suppressing stellar activity.
Quiescent galaxies (QGs) are known to host little to no star formation, as quenching processes transform their interstellar medium (ISM) suppressing stellar activity. While their stellar component has been extensively studied out to $z \sim 7$ (\citealt{weibel25}), its connection to quenching and late-stage cold ISM evolution remains poorly understood. The long-standing paradigm of QGs being deprived of cold ISM is now challenged by growing evidence of significant molecular gas and dust up to $z\sim3$. These seminal works push the limit of faint ISM detection via stacking (\citealt{man2016}, \citealt{gobat2018unexpectedly}; \citealt{magdis2021interstellar}), few individual ALMA detections (\citealt{morishita22}; \citealt{lee2023high}; \citealt{spilker25}), or via reddened attenuation profiles observed with the James Webb Space Telescope (JWST; \citealt{setton24}; \citealt{siegel2025}).
%Although the stellar component of QGs has been extensively studied out to $z \sim 7$ (\citealt{weibel25}), the evolution of their cold ISM and its connection to quenching remains poorly understood due to its faint nature. The traditional view of QGs being devoid of cold ISM has been challenged by growing evidence of significant molecular gas and dust up to $z\sim3$ inferred via stacking (\citealt{man2016}, \citealt{gobat2018unexpectedly}; \citealt{magdis2021interstellar}), a handful of individual ALMA detections (\citealt{morishita22}; \citealt{lee2023high}; \citealt{spilker25}), and indirectly, via reddened attenuation profiles observed with the James Webb Space Telescope (JWST; \citealt{setton24}; \citealt{siegel2025}).

Interpreting such measurements is challenging because wide range of methods often compensates for the lack of individual detections. Estimates of cold molecular gas masses ($M_{\rm H_2}$) make use of CO and [CII] signatures, %to be converted into $\rm H_2$ through the $\alpha_{\rm CO}$ (see \citealt{Bolatto2013review} for an extensive review) conversion factor, generally dependent on both metallicity and redshift (\citealt{sandstrom2013co}) and poorly known for quiescent systems. In particular, such techniques are 
and are mostly biased to recently quenched or ultra-massive systems (\citealt{belli21}; \citealt{williams21}; \citealt{bezanson21}; \citealt{wu23}; \citealt{zanella23}; \citealt{suess25}; \citealt{Umehata_2025}). For more evolved QGs, such constraints are rare \citep{spilker2018molecular}, and typically rely on the use of dust continuum (\citealt{donevski23}) from stacking \citep{gobat2018unexpectedly, magdis2021interstellar, blanquez2023gas, adscheid25}, since individual ALMA dust continuum detections are limited to only a handful of QGs at $z\!\sim\!1-3$ \citep{whitaker21b, morishita22, lee2023high}. 
%and are mostly biased to recently quenched or ultra-massive systems (\citealt{belli21}; \citealt{williams21}; \citealt{bezanson21}; \citealt{wu23}; \citealt{zanella23}; \citealt{suess25}; \citealt{Umehata_2025}). For more evolved QGs, constraints are rare \citep{spilker2018molecular}, and typically rely on use of dust continuum (\citealt{donevski23}) often from stacking \citep{gobat2018unexpectedly, magdis2021interstellar, blanquez2023gas, adscheid25}, since individual ALMA dust-continuum studies usually yield non-detections, with secure detections limited to only a handful of QGs at $z\!\sim\!1-3$ \citep{whitaker21b, morishita22, lee2023high}. 

Even when detected, dust traces $M_{\rm H_2}$ only via assumed dust-to-gas ratio ($\delta_{\rm DGR} = \rm M_{dust}/M_{gas}\!\sim\!1/100$), calibrated on star-forming galaxies (SFGs; \citealt{magdis2012evolving, scoville16}) and often treated as redshift independent \citep{popping2023dust}. Using this canonical conversion yields uncertain dust-based $\rm H_2$ fractions ($\sim1\%$–$20\%$; \citealt{gobat2022uncertain}). Together with the limitations of stacking, this challenges our understanding of processes acting on the cold ISM in individual QGs. 

%If dust measurements are already challenging, using them to obtain cold gas is an additional challenge, as the process typically relies the assumption of a constant conversion factor, the dust-to-gas mass ratio ($\delta_{\rm DGR}=M_{\rm dust}/M_{\rm H_2} \sim 1/100$). This fraction is calibrated on SFGs (e.g., \citealt{scoville16}) and it is observed to evolve with metallicity (\citealt{magdis2012evolving}) but not redshift (\citealt{popping17}). However, due to the lack of statistics the $\delta_{\rm DGR}$ is not constrained in QGs and the canonical value is used. Under this strong assumption, dust-based studies yield very uncertain gas fractions ($f_{\rm H_2}=M_{\rm H_2}/M_{\star}$) from $1\%$ to $\sim20\%$ \citep{gobat2022uncertain}.

%Dust and gas are typically coupled in star-forming galaxies (SFGs), but this link is poorly understood in QGs. Stacking cannot reveal the physical mechanisms acting on cold ISM in individual QGs, while inferring $\rm H_2$ mass from dust continuum relies on a constant dust-to-gas mass ratio ($\delta_{\rm DGR}=M_{\rm dust}/M_{\rm H_2} \sim 1/100$), calibrated on SFGs (e.g., \citealt{scoville16}) due to absent statistics for QGs. Under this strong assumption, dust-based studies yield very uncertain gas fractions ($f_{\rm H_2}=M_{\rm H_2}/M_{\star}$) from $1\%$ to $\sim20\%$ \citep{gobat2022uncertain}.% which, together with CO observations, yield $f_{\rm H_2}$ from $1\%$ to $20\%$ \citep{deugenio23, and references therain} 

Similar disagreement arises in interpreting the dust fraction ($f_{\rm dust} = M_{\rm dust}/M_{\star}$) for which early dust studies find a strong anti-correlation with stellar paopulation age, suggesting rapid dust removal within $\sim$150–250 Myr (e.g., \citealt{li19}), mirroring a fast ${\rm H_2}$ depletion during quenching (\citealt{whitaker21b}). However, this uniform picture has been questioned by works pointing to possible dust reformation in QGs (\citealt{donevski23}) and longer removal timescales ($>1$–2 Gyr; \citealt{lee2023high}; \citealt{michalowski23}). Theoretical studies based on \texttt{SIMBA} cosmological simulation (\citealt{simba2019}) further suggest a complex interplay of molecular gas and dust, producing a broad range of $\delta_{\rm DGR}$ in QGs (\citealt{whitaker21a}; \citealt{lorenzon25}). Whether this diversity reflects true physical complexity or stems from methodological biases in probing ISM in QGs remains an open question (\citealt{gobat2022uncertain}). 

Disentangling the impact of quenching pathways from ISM evolution in QGs requires direct, independent measurements of both dust and gas masses, i.e., their $\delta_{\rm DGR}$. %Attempts in this regards have been performed at high-$z$, where ISM content is observed to increase for SFGs (\citealt{tacconi2018phibss}) and might have a similar trend for QGs. %At intermediate redshift, \cite{donevski23} provides the largest sample of QGs with estimation of $\rm f_{\rm dust}$ using Herschel and Spitzer constraints, but lacking lower frequencies information. 
%Known attempts in this regard focus on high-$z$ as studies of SFGs found increase in ISM content from $z\sim0$ to $z\sim2$ \citep{tacconi2018phibss} and may show similar trends in QGs, but the scenry is more scarse at intermediate-$z$. 
Efforts so far have focused on high-$z$, where SFGs show an increase in ISM content from $z\!\sim\!0$ to $z\!\sim\!2$ \citep{tacconi2018phibss}, a trend QGs may share \citep{magdis2021interstellar}. %At intermediate redshift, the landscape remains far less explored. 
A first systematic study of five QGs at $z\!\sim\!1$ reported low but diverse values ($\delta_{\rm DGR}<1/300-1/1200$; \citealt{spilker25}). However, the limited sample size and bias towards the most massive and ${\rm H_2}$-rich QGs leave open the key question of whether molecular gas and dust evolve similarly after quenching, and whether this evolution is solely driven by their stellar properties such as age and mass. 
%As ISM content is observed to increase with redshift in SFGs (\citealt{tacconi2018phibss}), the same can be envisioned for QGs, leading most ISM studies in QGs to be at higher-z. Currently, the largest    

In this Letter, we present deep ALMA observations of 17 homogeneously selected QGs at $z\sim0.4$, simultaneously probing their cold dust and molecular gas. By jointly measuring dust and $\rm H_2$ masses, we aim to provide the first unbiased view of their co-evolution in the late phases of galaxy evolution. Throughout the paper we assume a \citealt{planck2018} cosmology.

\section{Observations of molecular gas and dust in QGs} \label{ALMA observations}

\subsection{Targets}
\begin{figure*}
    \includegraphics[width=0.99\textwidth]{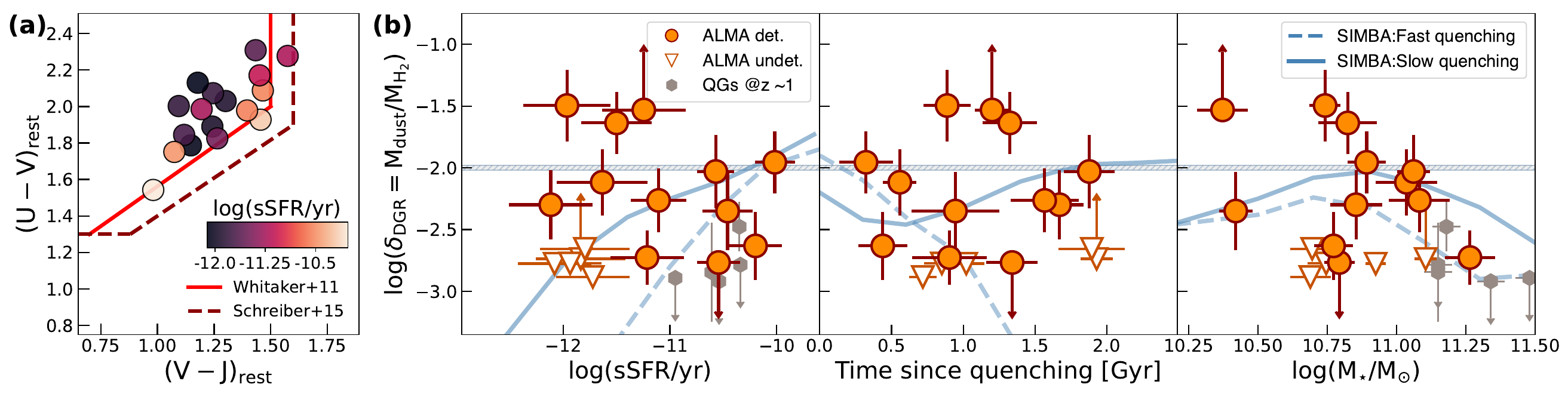}
    \centering
    \vspace{-0.25cm}
\caption{(a) Selected QGs in the UVJ plane, color-coded by $\rm sSFR$. 
(b) Evolution of $\delta_{\mathrm{DGR}}$ with $\rm sSFR$, $t_{\rm q}$, and $M_{\star}$. Large circles mark QGs detected with ALMA (10 in both CO and dust, one in either CO or dust); triangles show non-detections in both CO and Band~6 dust continuum. Small points are literature QGs at $z\sim1$ \citep{spilker25}. 
For all symbols, the lower or upper limits of $\delta_{\rm DGR}$ are denoted with arrows. Solid lines show fast and slow quenching tracks for $z\sim0.4$ QGs from \texttt{SIMBA} \citep{lorenzon25}. The horizontal line marks the canonical $\delta_{\rm DGR}\sim1/100$.}
    \label{fig:Fig1a}
\end{figure*}

Our ALMA Band~6 observations target 17 QGs in the COSMOS field drawn from a statistical parent sample of $\sim 500$ QGs combining deep, medium resolution spectroscopy (hCOSMOS; \citealt{damjanov18}) with $>$15 photometric bands from homogeneously calibrated catalogues (HELP; \citealt{shirley2019}). We refer to \citealt{donevski23} (hereafter \hyperref[d23]{D23}\label{d23}) for a detailed description of the sample. The 17 targeted QGs are selected to lie within a narrow redshift range ($0.33 < z < 0.43$; median of $z\!=\!0.36$),\footnote{This minimizes redshift-driven variation in $\mathrm{H_2}$ mass evolution.} and to be old (stellar population ages $>$3–10 Gyr) and massive ($10.4 < \log(M_{\star}/M_{\odot}) < 11.25$). The galaxies are required to simultaneously satisfy multiple criteria for quiescence: prominent \SI{4000}{\angstrom} break ($\mathrm{D_{n}4000 > 1.5}$), no $\mathrm{H_{\beta}}$ detection from deep optical spectroscopy, and low specific star formation rate (sSFR), $>0.6$ dex below the \cite{speagle2014highly} main sequence. All targets also satisfy the UVJ color criteria for quiescence (\citealt{schreiber15}; see Fig.~\ref{fig:Fig1a}a). Among them, 11/17 show tentative ($\sim$2–3$\sigma$) detections in de-blended, low-resolution \textit{Herschel} PACS and/or SPIRE maps (\hyperref[d23]{D23}), while the remaining 6 yield non-detections.% (see Table~\ref{apptable:sourcelist}). %These are listed in Table~\ref{apptable:sourcelist} as hCOS-d10, d12, d13, d15, d16, and d17. 

\subsection{ALMA Band 6 observations, data reduction and analysis}
We use data from ALMA project 2024.1.00814.S (PI: Lorenzon), obtained in multiple observing sessions between October 4 and December 1, 2024 (Cycle 11). Observations were conducted with the $12\:\rm m$ array in the C-3 configuration, with varying on-source integration time (30-95 minutes) totalling $26.4\:\rm h$. Band~6 was used to cover both the CO(3--2) line transition ($\nu_{\rm rest} = 345.796\:\rm GHz$) and the dust continuum. Integration times were set to reach a signal-to-noise ratio $\ge$5 at continuum. For the 6 PACS/SPIRE-undetected targets, CO emission instead set the limiting factor, assuming likely undetection even with deeper exposures and resulting in $\sim 2\times$ higher noise levels.

%The 6 targets with no PACS and/or SPIRE map detections were observed with roughly half the integration time, due to the expected low dust content and a presumed non-detection even under deeper exposures.

We use \texttt{CASA} \citep{McMullin2007casa} version 6.6.1.17 for both calibration (pipeline 2024.1.0.8) and imaging, conducted with the \texttt{tclean} task with per-source optimization. Continuum maps were extracted with multifrequency synthesis on spectral windows with no line features. Continuum-subtracted spectral cubes on the image plane were used to produce CO moment-zero maps. In both procedures we use a natural weighting scheme to maximize the recovery of large faint structures. 
We opted for untapered images to avoid contamination from nearby bright sources (4/17 cases; Fig.~\ref{appfig:ALMA_cut}) and to keep flux extraction homogeneous, having verified that a heavy taper ($2^{\prime\prime}$) only slightly raises the measured continuum fluxes.
%Since a UV taper of $4^{\prime\prime}$ (adequate for the largest sources) correspond to an increase in the measured continuum fluxes of $\lesssim 10\:\rm \mu Jy$ we opted for no taper at all to avoid contamination from bright nearby sources (4/17 cases; see Fig.~\ref{appfig:ALMA_cut}) while keeping a homogeneous flux extraction. 
The resulting beam sizes of $\sim0.8^{\prime\prime} - 0.9^{\prime\prime}$ allow for resolving most sources over 3–4 beams. We reach deep background levels of $\rm rms = 9\text{–}15 \:\mu Jy/beam$ in the continuum and $7\text{–}20 \:\rm mJy\,beam^{-1}\,km\,s^{-1}$ for the CO line. Fluxes are extracted using \texttt{Photutils} \citep{larry_bradley_2024_13989456} Kron apertures, selecting regions above $2\times \rm rms$ and applying de-blending in the case of close continuum emissions. Out of 17 targets, 10 show $>3\sigma$ detections in both dust and CO(3--2), one only in continuum, and one only in CO (see Fig.~\ref{appfig:ALMA_cut}). Among the remaining 5 QGs undetected with ALMA in both tracers, 4 lack archival IR data, while one has Spitzer MIPS and PACS 100 $\mu$m fluxes. Thus, both ALMA-detected and ALMA-undetected QGs include cases with and without archival IR detections. % In figures (Fig.\ref{fig:Fig1a}, \ref{fig:Fig3}) we show their $\delta_{\rm DGR}$ estimates from continuum and CO upper limits, though these are not listed in Table\ref{apptable:sourcelist} due to the absence of firm constraints. One ALMA-undetected source has Spitzer/MIPS and PACS 100$\mu$m fluxes, to which we add our 1 mm upper limit and yield an $M_{\rm dust}$ estimate and thus an upper limit on $\delta_{\rm DGR}$ (shown as a triangle with arrow)\footnote{Remarkably, of the 6 targets with no archival PACS/SPIRE detections, only 4 (d17, d16, d15, d12) remain fully undetected with ALMA.}.

Overall, our sample provides the largest dataset of QGs with joint dust and $\rm H_2$ measurements, tripling known samples beyond the local Universe. We stack the five ALMA–undetected QGs but find no signal, with $1\sigma$ sensitivities reaching  5 $\mu$Jy for dust continuum, and $10 \:\rm mJy\,beam^{-1}\,km\,s^{-1}$ for CO(3--2). 

\subsection{Physical properties}
To estimate the physical properties of QGs we use the \texttt{v2025 CIGALE} code (\citealt{cigale}) and perform spectral energy distribution (SED) fitting following the methodology from \hyperref[d23]{D23}. An extended description of our SED fitting is provided in \autoref{appendixA}. Briefly, we model the stellar emission using \cite{bruzual03}. We apply a flexible star-formation history (SFH) combining delayed and quenched components, and a two-component attenuation law with age-dependent reddening \citep{charlott00}. For dust emission, we employ \citet{dl07} library. Our dataset has broad optical-to-NIR coverage (15+ bands) and up to five mid-IR to sub-mm points (detections or upper limits) from \hyperref[d23]{D23}, supplemented with our ALMA data and, when available, JWST NIRCam (14/17 sources) and MIRI (9/17 sources) fluxes \citep{cosmos25}. We find good agreement between our fiducial $M_{\rm dust}$ to estimates from alternative methods (\autoref{appendixB}).

We infer $M_{\rm H_2}$ by using velocity-integrated fluxes from the moment-zero map, and applying the standard procedure (\citealt{solomon05}). We assume a line ratio of $R_{31} = L^{\prime}_{\rm CO(3-2)} / L^{\prime}_{\rm CO(1-0)} = 0.5$, and a constant conversion factor $\alpha_{\rm CO} = 4.36 \:\rm M_{\odot}(\rm K\:km\:s^{-1} pc^{-2})$ consistent with similar works (e.g., \citealt{smercina22}, \citealt{Umehata_2025}). 
%, appropriate for post-starbursts and QGs (e.g., \citealt{smercina22}, \citealt{Umehata_2025}). 

Together, this allows us to directly constrain $\delta_{\rm DGR}$ for the 10 QGs detected in both CO and dust, and to set lower and upper limits for the CO-only and dust-only detections, respectively. For the ALMA-undetected source with archival IR data, we combine its Spitzer/MIPS and PACS 100$\mu$m fluxes with our 1 mm flux upper limit to estimate $M_{\rm dust}$ and thus an upper limit on $\delta_{\rm DGR}$. For the remaining 4 ALMA-non-detections, throughout the paper we show $\delta_{\rm DGR}$ resulting from continuum and CO upper limits\footnote{These are not listed in Table~\ref{apptable:sourcelist} due to the lack of firm constraints.}. The sources and their derived properties are summarized in \autoref{appendixC} and \autoref{apptable:sourcelist}.
%\footnote{Remarkably, of the 6 targets with no archival PACS/SPIRE detections, only 4 (d17, d16, d15, d12) remain fully undetected with ALMA.}.

\section{Direct probe of \texorpdfstring{$\delta_{\rm DGR}$}{delta\_DGR} in old QGs}
\label{DGR evolution}

With our sample, we directly access $\delta_{\rm DGR}$ which quantifies the fraction of the ISM mass in dust grains \citep{popping2023dust}. In Fig.~\ref{fig:Fig1a}b we illustrate a striking diversity in $\delta_{\rm DGR}$ across stellar properties, namely specific star-formation rate (sSFR), time since quenching ($t_{\rm q}$; defined as a lookback time to the truncation of star formation), and stellar mass ($M_{\star}$). The variation in $\delta_{\rm DGR}$ exceeds an order of magnitude, even excluding ALMA-undetected QGs. This large spread ($\delta_{\rm DGR}\sim1/700$ to $1/40$) deviates by $\sim8\times$ below and $\sim2.5\times$ above the canonical value of $\delta_{\rm DGR}\sim1/100$ regardless of how $M_{\rm dust}$ and $M_{\rm H_2}$ are estimated (see \autoref{appendixB}). This marks the first direct confirmation that $\delta_{\rm DGR}$ in QGs can exceed values typical for SFGs.  

The broad range of $\delta_{\rm DGR}$ suggests that QGs contain complex ISM despite their low sSFR. Notably, the left and mid-panel of Fig.~\ref{fig:Fig1a}b reveal that the highest $\delta_{\rm DGR}$ arise at low sSFRs, $\gtrsim1$ Gyr post-quenching, weakening the link to residual star formation. Interestingly, the presence of similarly massive, dust-attenuated QGs persisting for $>1\:\rm Gyr$ after quenching have recently been independently revealed by MIRI observations (Lisiecki et al., submitted). Although we do not see strong anti-correlation with stellar mass, QGs with $\delta_{\rm DGR} \gtrsim1/100$ are restricted to $M_{\star} < 10^{11}\,M_{\odot}$. This supports results from dust stacking studies at high-$z$ \citep{blanquez2023gas} and supports the idea that the most massive QGs limit $\delta_{\rm DGR}$ due to efficient dust destruction or heating in massive halos \citep{zheng2022rapidly, lorenzon25, spilker25}. Similar arguments apply to massive lensed QGs at $z\gtrsim1$-2 \citep{williams21, whitaker21b}, where continuum non-detections likely reflect rapid dust removal or destruction rather than insufficient ALMA depth. %This is shown in Fig. ~\ref{fig:Fig2}, by the diamond lying on the rapidly decaying exponential curve fitted to our dust non-detections.

% However, this does not resolve the tension with studies of $z\gtrsim2$ QGs favouring fully depleted dust in massive QGs (\citealt{williams21}, \citealt{whitaker21b}). 

In Fig.~\ref{fig:Fig1a}b, we show \texttt{SIMBA} simulation tracks for rapid and slow quenching modes, separated at a quenching time of $t_{\rm q}\sim 2\times 10^8\:\rm yr$ (\citealt{lorenzon25})\footnote{This cut was found in \citealt{lorenzon25} as a broad separation between the two peaks of the normalized quenching time distribution (see also \citealt{rodriguez2019mergers} and \citealt{zheng2022rapidly}). The redshift-dependent cut is applied here to $\rm z=0.36$}. These modes broadly (but not exclusively) reflect feedback efficiencies from active galactic nuclei (AGN), with long-term gas heating contributing post-quenching. Our QGs span a wide range of $t_{\rm q}$, suggesting influence from multiple quenching channels (e.g., \citealt{belli21}, \citealt{park2023rapid}). As seen in Fig.~\ref{fig:Fig1a}b, both ALMA-detected and undetected QGs do not align cleanly with fast or slow quenching tracks. This suggests that ISM and stellar properties in QGs are not solely determined by quenching mode but by a more complex interplay \citep{lorenzon25}. The QGs with larger $\delta_{\rm DGR}$, however, are broadly consistent with slow quenching, often lying above modeled tracks, suggesting that dusty QGs may be common among slowly quenching, intermediate-mass systems ($M_\star<10^{11}M_\odot$; \citealt{lorenzon25}).

%residing along the slow quenching track in Fig.~\ref{fig:Fig1a}a, but being closer to the fast track in Fig.~\ref{fig:Fig1a}b and Fig.~\ref{fig:Fig1a}c. 

%This points to ISM and stellar properties in QGs to not exclusively depend on the modality of quenching, but rather on a more complex interplay (\citealt{lorenzon25}). Galaxies with larger $\delta_{\rm DGR}$, however, tend to be more broadly compatible with the slow quenching scenario and, in most cases, to be far above the spread of simulated tracks. This supports the possibility that such dusty QGs are common within slowly quenching galaxies of intermediate masses ($<10^{11}\:M_{\odot}$).

%ALMA-undetected QGs favour the fast quenching channel, while detected QGs span both modes with broad $\delta_{\rm DGR}$ and a mild preference for slow quenching. Our data generally follow simulated tracks, but dust-rich outliers stand out. As shown in \citealt{lorenzon25} (see also \citealt{akins2022quenching}), the emergence of such QGs is hard to explain by quenching mode alone. Though possible in both modes, such dusty QGs are predicted to be common within slowly quenching galaxies of intermediate masses ($<10^{11}\:M_{\odot}$). Our data support this possibility.
In SFGs, $\delta_{\rm DGR}$ variations are often linked to gas-metallicity (e.g., \citealt{de2019systematic}), but this is unlikely for old QGs, which occupy a narrow gas-phase metallicity range (e.g., \citealt{pistis24}). We note that $\delta_{\rm DGR}$ in this study concerns $\rm H_2$ gas, ignoring atomic H\textsc{I}. Known studies on early-type galaxies and QGs in the local Universe confirm they can host significant H\textsc{I} reservoirs (up to $\gtrsim 10^{10} M_{\odot}$; \citealt{michalowski23}, \citealt{ellison25}). However, %to reconcil the highest $\delta_{\rm DGR}$ in our sample with the $\delta_{\rm DGR}\sim1/100$ would require H\,\textsc{i} masses up to $3\times10^9\:M_\odot$, but 
none of our QGs is individually detected in the MeerKAT-HI survey data with mass limit down to $\sim 5\times10^9\:M_\odot$ \citep{bianchetti25}. Applying our observational selection criteria in \texttt{SIMBA}, we find  only $13.05\pm1.08\:\%$ of QGs above this mass limit. Furthermore, studies at $z\sim0$ reveal decrease in H\textsc{I}-detection fraction towards massive QGs ($\log(M_{\star}/M_{\odot})\gtrsim10.5$; i.e., \citealt{Guo2021}, \citealt{ellison25}). We thus find it unlikely that our QGs are exceptionally H\textsc{I}-rich, and assume that the spread in $\delta_{\rm DGR}$ rather reflects diverse timescales for dust and $H_2$ gas evolution, which we explore in the following sections.

%Such modest fraction of $M_{\rm H\:I}$ is consistent with extrapolation from low-$z$ studies finding %also predict similarly modest $M_{\rm H\:I}$ for QGs of similar stellar masses as ours (\citealt{Guo2021}).

%%We thus rule out that those are as H\,\textsc{i} bright targets as those “HI-rich” QGs found in the local Universe. 
%Such large values are in contrast with what measured for low-$z$ QGs in haloes of masses lower than $10^{13}\:\rm M_{\odot}$ (\citealt{Guo2021}).   

%aHowever, %to reconcil the highest $\delta_{\rm DGR}$ in our sample with the $\delta_{\rm DGR}\sim1/100$ would require H\,\textsc{i} masses up to $3\times10^9\:M_\odot$, as high as the total $M_{\rm H\rm I}$ inferred from stacking QGs at $z\sim0.4$ in the COSMOS field \citep{bianchetti25}. We find it unlikely that our QGs are exceptionally H\,\textsc{i}-rich, and assume that the spread in $\delta_{\rm DGR}$ rather reflects diverse timescales for dust and $H_2$ gas evolution, which we explore in the following sections.

\subsection{Evolution of dust and $H_2$ gas fraction with stellar population age}
\label{fdust}
To explore the drivers of $\delta_{\rm DGR}$ variations in our QGs, we examine the age evolution of dust fraction ($f_{\rm dust} = M_{\rm dust}/M_{\star}$) and gas fraction ($f_{\rm H_2} = M_{\rm H_2}/M_{\star}$). Our sample contains older systems (mass-weighted ages $\gtrsim$~3 Gyr), minimizing contamination from recently quenched objects. Combining it with the depth of our ALMA we can access previously unexplored regions of the $f_{\rm dust}$–age plane in old QGs ($>5$–10 Gyr), providing new constraints on dust depletion timescales.

\begin{figure}[h!]
    \includegraphics[width=\columnwidth]{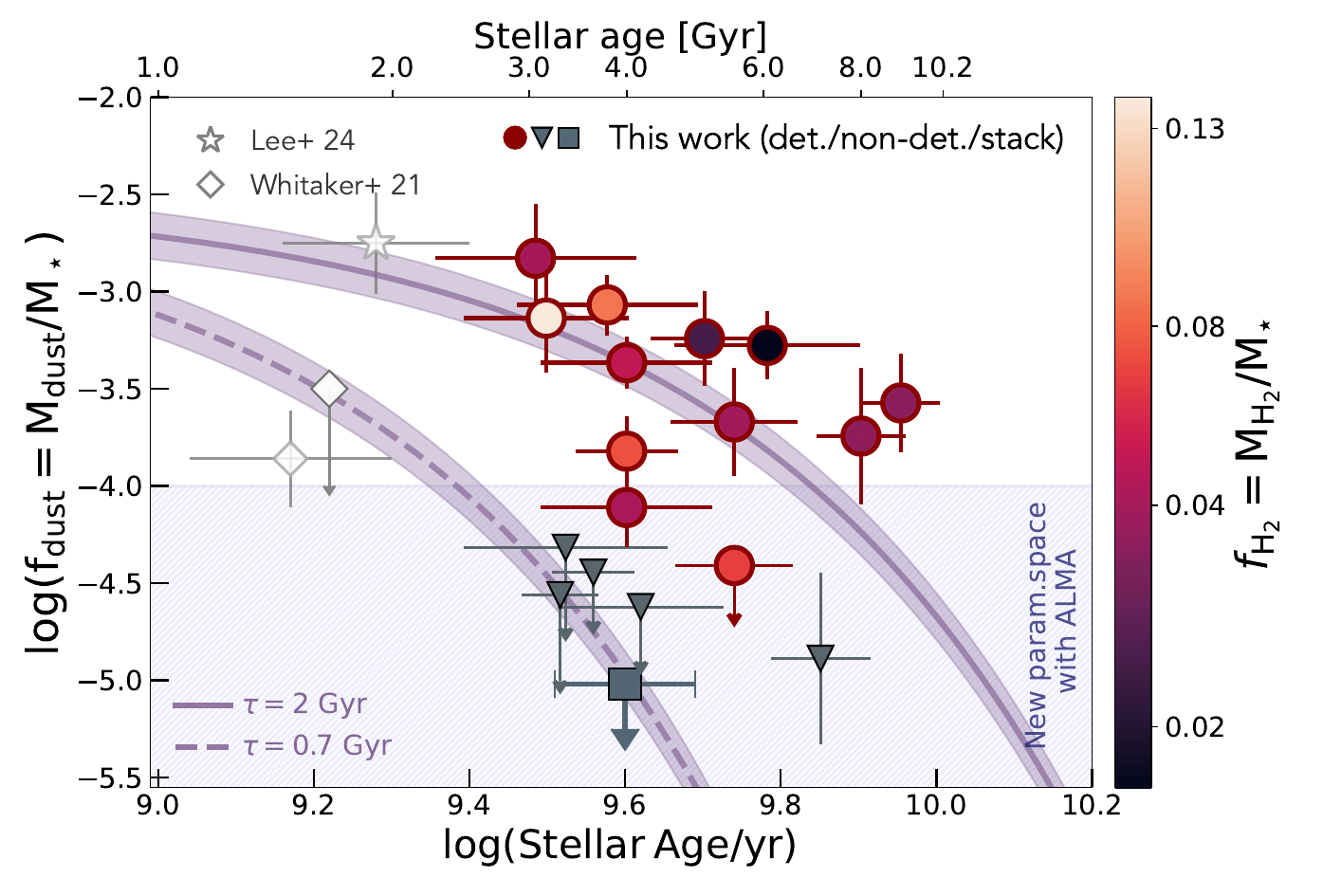}
    \centering
    \vspace{-0.25cm}
\caption{Relation between $f_{\mathrm{dust}}$ and mass-weighted stellar age, color-coded by $\rm H_2$ fraction. As in Fig.~\ref{fig:Fig1a}, circles and triangles mark ALMA-detected and non-detected QGs, respectively, with arrows denoting upper limits. The single triangle with an $f_{\rm dust}$ constraint corresponds to a PACS-detected but ALMA-undetected QG. The gray square shows the stacked non-detections in Band~6 dust continuum. Curves trace exponential tracks for fast and slow dust removal from \cite{lorenzon25}. White symbols indicate literature QGs at $z\sim2$. The shaded region highlights the previously unexplored $f_{\mathrm{dust}}$–age space of old QGs.}
    \label{fig:Fig2}
\end{figure}

We probe models with exponential decline ($Ae^{-t/\tau_{\rm dust}}$), adopting short ($\tau_{\rm dust}\sim 0.7\:\rm Gyr$) and long ($\tau_{\rm dust}\sim 2 \: \rm Gyr$) removal timescales normalised to the typical $f_{\rm dust}$ of main-sequence SFGs at $z<1$ \citep{donevski20}. These timescales reflect distinct post-quenching evolutionary paths found in \texttt{SIMBA} \citep{lorenzon25}, with shorter $\tau_{\rm dust}$ corresponding to efficient dust destruction via thermal sputtering and heating, typically AGN-driven (i.e. \citealt{hirashita2017dust}). Notably, our QGs as well as dust-studied ones at $z\sim2$ \citep{whitaker21b, lee2023high} do not align with a single track in $f_{\rm dust}$ with age (Fig.~\ref{fig:Fig2}), imposing diverse dust depletion (or removal) timescales. 

While early studies of post-starburst and elliptical galaxies report parallel dust and gas declines over similar timescales (e.g., $<400\:\rm Myr$ for post-starbursts and up to $2.5\:\rm Gyr$ for ellipticals; \citealt{li19, bezanson21, michalowski23}), our findings challenge this view. First, an increase in $f_{\rm dust}$ at fixed stellar age is not fully mirrored by $f_{\rm H_2}$. Second, sustained high $f_{\rm dust}$ towards older ages in gas-poor QGs seems incompatible with passive, age-driven depletion. Longer dust depletion timescales may arise if dust grains are sufficiently large to slow destruction by sputtering, as shown by \cite{Nanni2025} and with SIMBA in \cite{lorenzon25}, which invoke replenishment either via minor mergers or internally, via re-growth on metals over $\lesssim$150 Myr timescales, flattening the $f_{\rm dust}$–age evolution. Therefore, our findings support \texttt{SIMBA} prediction that, \textit{for at least some ISM-rich QGs}, dust and cold gas may follow divergent paths after quenching.

\subsection{Tracing QGs post-quenching routes with $H_2$}
\label{fgas}
\begin{figure*}
    \centering
    \includegraphics[width=0.9\textwidth]{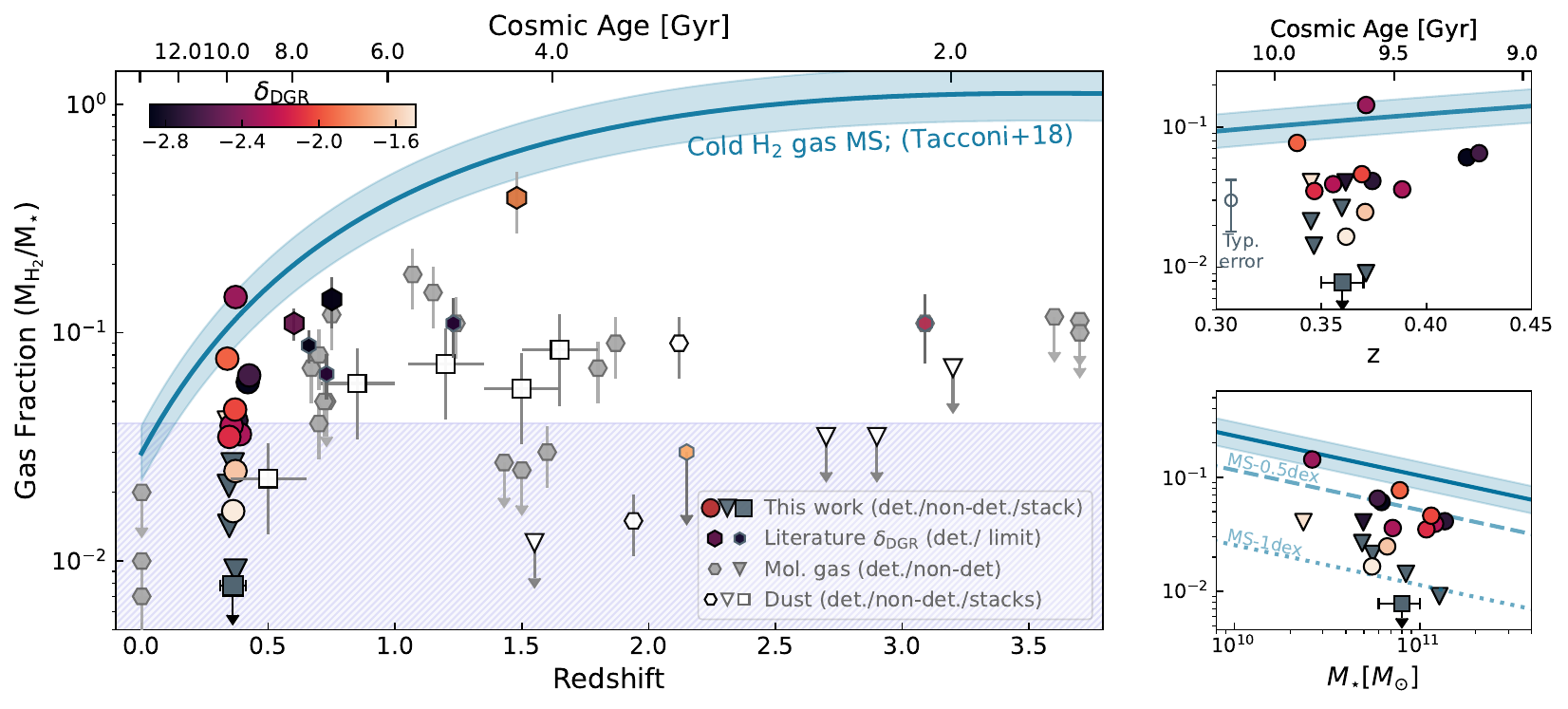}
    \caption{$\rm H_2$ gas fraction vs. redshift for our sample and literature QGs, with detectability (CO- or dust-based) noted in the legend. The points include QGs at $z\sim1$ (\citealt{belli21}, \citealt{woodrum2022molecular}, \citealt{bezanson21}), $z\sim2\text{–}3$ (\citealt{whitaker21b}, \citealt{morishita22}, \citealt{williams21}, \citealt{suzuki2022low}, \citealt{Umehata_2025}), and stacks (\citealt{magdis2021interstellar}, \citealt{blanquez2023gas}). Coloured symbols show $\delta_{\rm DGR}$, directly estimated only for our sample, two QGs at $z\sim0.7$ (\citealt{spilker25}) and QG at $z\sim1.5$ \citep{hayashi18}, with others showing limits. The main-sequence $\rm H_2$-scaling relation \citep{tacconi2018phibss} is shown in blue. The shaded area shows the new parameter space of $f_{\rm H_2}$ opened for individual detections. Right panel zooms-in on our QGs relative to the main-sequence relation at their $z$ and $M_{\star}$. The typical error on $f_{\rm H_2}$ for our QGs is shown in the upper right sub-panel.}
    \label{fig:Fig3}
\end{figure*}

By reaching low $\rm H_2$ fractions of $f_{\rm H_2}\!\lesssim\!4\%$ in individual, unlensed QGs (Fig.~\ref{fig:Fig3}), we open a new window onto lowest levels of residual gas, allowing us to reassess how well $f_{\rm H_2}$ traces the final stages of galaxy evolution. The CO-detected QGs have $1.7\% \lesssim f_{\rm H_2}\lesssim 14\%$ (median $4.1^{+1.2}_{-0.9}\%$), reaching $0.6\%$ in stacked non-detections. These uniformly low $\rm H_2$ place our QGs $4$–$10\times$ below the main-sequence $\rm H_2$ scaling of \citet{tacconi2018phibss}, consistent with offsets seen in QGs at $z\sim1$ \citep{williams21, woodrum2022molecular, suess25}.

Theoretical suggestion that CO-detected QGs can arise from both slow and fast quenching modes (as shown in Fig.\ref{fig:Fig1a}) is empirically supported by their gas depletion timescales ($\tau_{\rm dep} \propto M_{\rm H_2}/\rm SFR$). Combining $M_{\rm H_2}$ with SED-based SFRs yields $\tau_{\rm dep}$ from 0.4 to 48 Gyr. We find recently quenched QGs ($t_{\rm q}<$1 Gyr) having shorter $\tau_{\rm dep}$ ($\sim$0.4–0.8 Gyr) then older QGs who average around $\tau_{\rm dep}\sim$10 Gyr, longer than their $t_{\rm q}$. Such QGs are consistent with inefficient star formation and quenching via slow suppression rather than gas expulsion, and may require additional gas removal to fully quench. This aligns with findings for $z\sim1$ QGs, where diverse pathways imply $\rm H_2$ depletion alone cannot explain quenching \citep[e.g.,][]{belli21, wu23}. 

The $f_{\rm H_2}$ evolution diverges not only from the stellar component but also from dust, as Fig.\ref{fig:Fig3} shows no clear correlation with $\delta_{\rm DGR}$ across $M_{\star}$. We find a dispersion of $\sim$0.8 dex ($\sim$1.4 dex with stacked QGs) in $f_{\rm H_2}$, smaller than $\sim$1.3 dex (up to $\sim$2 dex including stacks) in $f_{\rm dust}$ for the same sample (Section~\ref{fdust}). This indicates that variations in dust content dominate the scatter in $\delta_{\rm DGR}$ at fixed $M_{\star}$. Strikingly, Fig.~\ref{fig:Fig3} unveils that in the newly accessed regime of very low $\rm H_2$ fractions, two QGs emerge with $\log\delta_{\rm DGR}\!\gtrsim\!-\!2$, enhanced above typical SFG values. Our most massive QGs ($>10^{11}M_{\odot}$) remain relatively gas-rich, $\sim\!5\times$ below the SFG gas-scaling from \citet{tacconi2018phibss}, but all exhibit $\delta_{\rm DGR} \lesssim 1/100$, falling below the standard value. This aligns with results from \cite{spilker25}, whose two $z\!\sim\!1$ QGs have low $\delta_{\rm DGR}=1/300$ and $1/750$. 

In \texttt{SIMBA}, most QGs of $\log(\rm M_{\star}/M_{\odot})>10.8$ undergo fast AGN-driven quenching \citep{zheng2022rapidly}, but whether this leads to full ISM removal depends on balance between multiple processes acting on gas and dust (i.e., dust re-growth, \hyperref[d23]{D23}). The high-$z$ QGs, small color-coded symbols in Fig.\ref{fig:Fig3}, illustrate this complexity: the $z\!\sim\!2$ QG from \citet{morishita22} has lower limit of $\delta_{\rm DGR} \gtrsim 1/60$ despite AGN outflow and lack of $\rm H_2$ gas\footnote{This results from non-detection of $[\mathrm{C\textsc{I}}]{}^{3}\!P_{2} \rightarrow {}^{3}\!P_{1}$ line emission.}; conversely, the dust-undetected but $\rm H_2$-rich QGs from \citet{Umehata_2025} and \cite{spilker25} contain suppressed $\delta_{\rm DGR}\lesssim$1/500–1/1000. Only $z\!\sim\!1.5$ QG (\citealt{hayashi18}) shows typical $\delta_{\rm DGR}$, but it is a distant, cool-core cluster object, likely an outlier to the general QG population. These cases suggest two key implications: (1) Dust is an unreliable tracer of $\rm H_2$ mass in QGs; (2) Non-uniform ISM evolution in the quenched phase may be common for QGs across redshifts, not merely an artifact of observational methods. This challenges the idea that quenching mode is the sole driver of ISM diversity, supporting models where feedbacks and dust reformation jointly shape the late ISM fate \citep{hirashita2017dust, lorenzon25}.

We further explore quenching modes and post-quenching evolution in two accompanying works: one provides an in-depth analysis of the gas and stellar structure and kinematics in a few extreme cases driving the $\delta_{\rm DGR}$ spread (Donevski et al., in prep.),  while the other provides a statistical treatment accounting for morphological diversity and AGN radio signatures in our QGs, and characterizes their impact on dust-gas evolution (Lorenzon et al., in prep.).

\section{Conclusions}
\label{Conclusions}
We present the first statistical sample of 17 massive QGs at $z\!\sim\!0.4$, probing their ISM with ALMA Band 6 dust continuum and CO(3–2) emission. For 12/17 systems we constrain key diagnostics ($\delta_{\rm DGR}$, $f_{\rm dust}$, $f_{\rm H_2}$) and link them to stellar properties, finding that:
\begin{itemize}
\item Dust-to-gas ratios in QGs span $\delta_{\rm DGR}\sim1/700$–$1/40$, departing $\sim8\times$ below to $\sim2.5\times$ above the canonical value of $\delta_{\rm DGR}\sim1/100$. Although most QGs align with slow-quenching mode tracks, diverse ISM across stellar mass, age, sSFR, and quenching timescale imply that quenching mode alone cannot explain the observed scatter.
\item The relation between dust fraction and stellar population age shows that some QGs undergo rapid dust decline (within $\tau\sim0.7$ Gyr), while others retain stable values ($\log f_{\rm dust} \gtrsim -3.5$) even towards the old ages ($>6$–$10$ Gyr) and $>1$ Gyr after quenching, supporting scenarios where dust reformation shapes the late ISM.
\item QGs show no clear relation between molecular gas and dust content across stellar mass, calling into question the reliability of dust as an $\rm H_2$ tracer.
\item Despite uniformly low $\rm {H_2}$ fraction (median $f_{\rm H_2}\sim 4.1^{+1.2}_{-0.9}\%$), the non-monotonic behaviour of $f_{\rm H_2}$, $f_{\rm dust}$, and stellar properties reveals diverse post-quenching pathways in QGs. Combined with $z\!\sim\!1$–3 studies, this suggests that cold ISM evolves through mechanisms beyond a simple decline of star-forming conditions.
\end{itemize}

Our study underscores the challenge of linking quenching modes to post-quenching ISM conditions, even with ALMA CO and dust detections. It motivates future model refinements to account for processes that decouple gas and dust to properly simulate ISM in QGs. To this end, a new parameter space of $f_{\rm H_2} < 4\%$ opened for individual QGs, highlights the need for JWST+ALMA studies of resolved ISM in QGs across cosmic time.

\begin{acknowledgments}
G.L and D.D acknowledge support from the NCN SONATA grant (UMO-2020/39/D/ST9/00720). D.D acknowledges support from the Polish National Agency for Academic Exchange (Bekker grant BPN /BEK/2024/1/00029/DEC/1). AWSM acknowledges the support of the Natural Sciences and Engineering Research Council of Canada (NSERC) via grant reference number RGPIN-2021-03046. S.B. is supported by ERC grant 101076080. J. is funded by the European Union (MSCA  EDUCADO, GA 101119830 and WIDERA ExGal-Twin, GA 101158446). A.N. and P.S. acknowledge support from the Narodowe Centrum Nauki (UMO2020/38/E/ST9/00077). This paper makes use of the following ALMA data: ADS/JAO.ALMA\#2024.1.00814.S. ALMA is a partnership of ESO (representing its member states), NSF (USA) and NINS (Japan), together with NRC (Canada), NSTC and ASIAA (Taiwan), and KASI (Republic of Korea), in cooperation with the Republic of Chile. The Joint ALMA Observatory is operated by ESO, AUI/NRAO and NAOJ. This research made use of Photutils, an Astropy package for detection and photometry of astronomical sources (Bradley et al. 2024). C.P. was supported through DL 57/2016 (P2460) from the ‘Departamento de Física, Faculdade de Ciências da Universidade de Lisboa’. C.P. acknowledges the suppor by Fundação para a Ciência e a Tecnologia (FCT) through the research grants UIDB/04434/2020 DOI: 10.54499/UIDB/04434/2020 and UIDP/04434/2020 DOI: 10.54499/UIDP/04434/2020. IS acknowledges fundings from the European Research Council (ERC) DistantDust (Grant No.101117541) and the Atracc\'{i}on de Talento Grant No.2022-T1/TIC-20472 of the Comunidad de Madrid, Spain. K.L acknowledges the support of the National Science Centre, Poland through the PRELUDIUM grant UMO-2023/49/N/ST9/00746.

\end{acknowledgments}

\bibliographystyle{aasjournal}
\bibliography{bibliography}

\twocolumngrid

\appendix
\section{Multiwavelength SED modeling and fitting of QGs}
\label{appendixA}
\subsection{SED fitting methodology}

We use the \texttt{CIGALE} code (\citealt{cigale}), v2025.0, and follow the approach of \hyperref[d23]{D23} to estimate the physical properties of our QGs. For the stellar component we use \citet{bruzual03} (BC03) stellar population synthesis with a \citet{chabrier03} initial mass function (IMF). We sample the stellar metallicity grid points closest to the mass–metallicity relation for each galaxy. We adopt a flexible SFH, modeling a delayed plus quenched component (\citealt{ciesla21}), given as:
\begin{equation}
\label{Eq.1}
\rm SFR=\begin{cases}
\mathit{t}\times e^{-t/\tau_{main}}, & \text{when $t\leq t_{\rm trunc}$}\\
r\times \rm SFR(\mathit{t}), &\text{when $t>t_{\rm trunc}$}
\end{cases}
\end{equation}
where $\tau_{main}$ is the e-folding time of the main stellar population, while $t_{\rm trunc}$ represents the time at which star formation is truncated, either instantaneously ($r_{\rm SFR}=0$) or partially ($0<r_{\rm SFR}<1$). The parameter $r$ is the ratio between SFRs after quenching and at quenching. We also test non-parametric SFHs by fitting the functional form to the center of the first five age binsm finding results consistent with our fiducial run. We use the \citet{charlott00} double power-law attenuation model (CF00) with age-dependent differential attenuation between young ($<10^7\:\rm yr$) and old stars. For dust SED we adopt the physically motivated \cite{dl07} library, proven efficient in modeling various galaxy types including QGs (\citealt{magdis2021interstellar}, \citealt{blanquez2023gas}). We fix the radiation parameter $U_{\rm max}=10^6$, emission slope ($\beta=2$), and sample different intensities $U_{\rm min}$. We let PAH fraction vary ($0.47<q_{\mathrm{PAH}}<4.6$) and limit illumination fraction $\gamma$ to $0.01$, reducing the chance to overestimate $M_{\rm dust}$. We define $L_{\rm IR}$ as the integral of the SED from 8–1000\,$\mu$m and derive $M_{\rm dust}$ by fitting and normalizing the IR photometry to the DL07 library. 

Our dataset includes rich sampling with at least 15 optical-to-NIR bands. At longer wavelengths it contains our ALMA Band 6 dust continuum fluxes (11 objects) and upper limits (6 objects). Where available, we complement ALMA photometry with mid-IR and/or far-IR fluxes estimated from de-blending deep \textit{Spitzer} MIPS and \textit{Herschel} PACS/SPIRE maps (\hyperref[d23]{D23}), and with JWST MIRI fluxes \citep{cosmos25}. Overall, 11/17 objects resulted in two to four photometric points in the mid-IR to sub-mm regime, ensuring reliable coverage of the Rayleigh-Jeans tail and proper normalisation of the dust SED, key for $M_{\rm dust}$ constraints. In our SED fitting we exploit the independent prior on $D_{\rm n}4000$ measured from optical spectra. As $D_{\rm n}4000$ is a reliable tracer of stellar age (e.g., \citealt{damjanov18}), we use the $D_{\rm n}4000$ option in CIGALE and iterate until measured and SED-derived $D_{\rm n}4000$ agree within $\pm 0.1$. This procedure ensures that modeling correctly matches stellar population ages, reducing dust–age degeneracies. For all physical quantities used throughout this study, we adopt Bayesian (probability-weighted average) values. Quality assessment reveals good fits for all modeled SEDs, with reduced $\chi^{2}$ ranging from 0.2 to 2.5, and median $\chi^{2}=1.02\pm0.33$.
 %We also inspect the CIGALE mock catalogues to confirm no significant systematics affect our dust estimates. Finally, comparing DL07-derived $M_{\rm dust}$ to optically thin RJ models calibrated on ALMA Band6 data (\citealp{scoville16}), shows excellent agreement within 0.2 dex (see Appendix C).

\subsection{Inspecting SED fitting systematics}

To assess the robustness of our SED fitting approach and quantify potential biases, we exploit a widely used method (e.g., \citealt{ciesla21}) for generating a simulated data set and analysing it using the same methodology applied to our observed galaxies. The primary goal is to evaluate how observational effects might influence the recovery of physical parameters through SED modeling. We use the "mock object" functionality within \texttt{CIGALE} to construct a synthetic catalogue, leveraging the best-fit SED model of each galaxy in our sample. This creates one artificial object per galaxy, for which the input physical parameters are known by construction. These models are then integrated through the same set of photometric filters as used in the observed data. To simulate observational uncertainties, we perturb the model fluxes by adding Gaussian noise, with $\sigma$ matching the measured uncertainty in each photometric band. The mock galaxies are then re-fitted using the same grid of physical models and priors as in the real sample, enabling a direct comparison between the recovered (output) and true (input) parameters. 

\begin{figure*}[tbh!]
    \centering
    \includegraphics[width=14.7cm,clip]{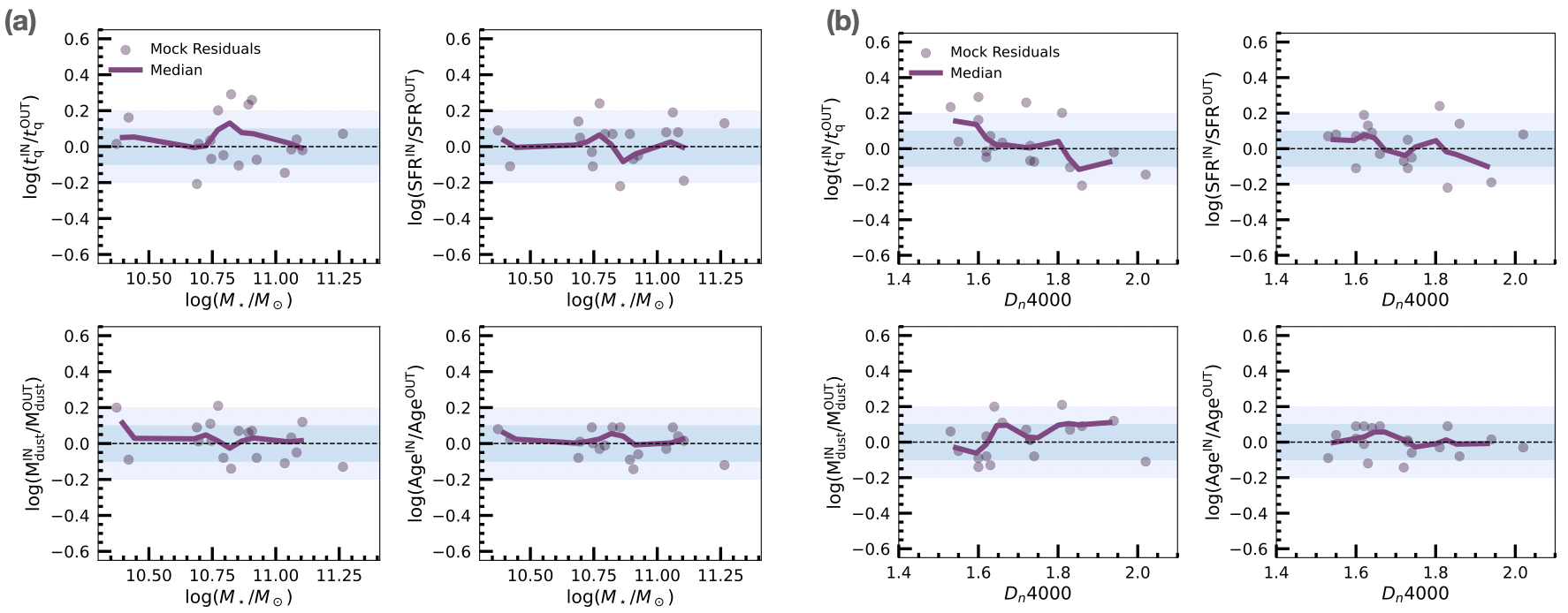}
    \caption{Results of our mock CIGALE analysis quantifying the offset between the mock input (“true”) and SED-fitting output (“observed”) parameters, as a function of $M_{\star}$ and independently measured $D_{\rm n}4000$. Dark violet lines show binned means, indicating good constraints with no systematic offsets.}
    \label{appfig:SED}
\end{figure*}

In \hyperref[appfig:SED]{Figure B.1}, we show the residuals between input and output parameters as a function of $M_{\star}$ and $D_{\rm n}4000$ index, measured from optical spectra. This test reveal a consistency between input and output values, as recovered trends for all physical parameters (namely, stellar population age, stellar mass, dust mass and time since quenching) closely follow the one-to-one relation, defined by a small mean offset and low scatter around zero value. In summary, this supports the reliability of the derived physical properties across our QG sample. %with $r^2$ values of 0.87 and 0.93, respectively. 

%To further investigate potential systematic effects, we examine the residuals (in log scale) between input and output parameters as a function of $M_{\star}$ and independently measured $D_{\rm n}4000$ index. A well-constrained parameter is defined by a small mean offset and low scatter around zero. The lack of significant trends with $D_{\rm n}4000$ confirms that our SED fitting is not biased with respect to the strength of the 4000\AA break, and supports the reliability of the derived physical properties across our QG sample.

\section{Comparison of different methods for deriving $M_{\rm dust}$}
\label{appendixB}

\begin{figure}[tbh!]
    \centering
    \includegraphics[width=11.33cm,clip]{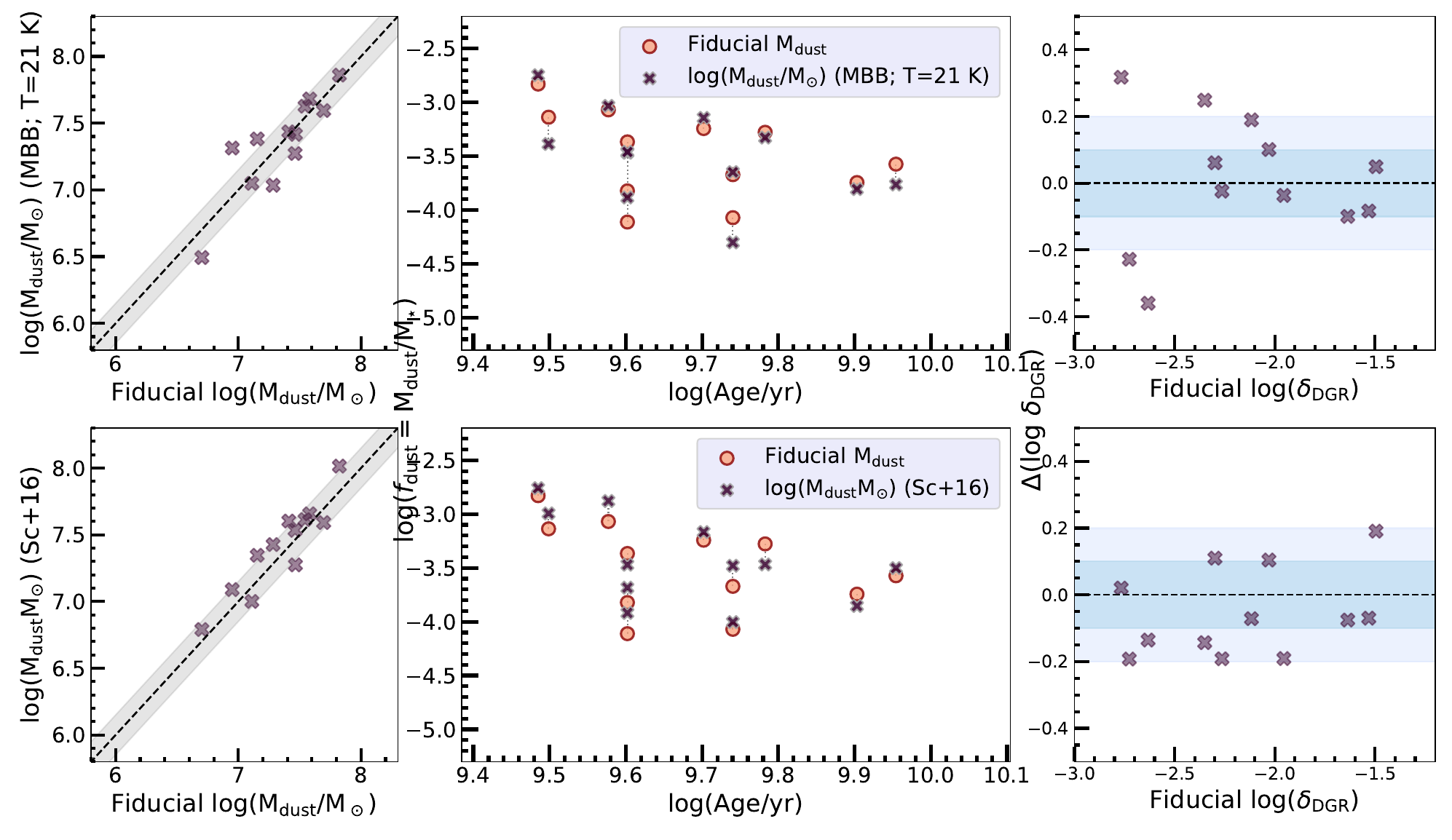}
    \caption{Comparison of fiducial dust properties with alternative dust-mass estimates (Upper: vs. modified blackbody fits; lower: vs. \cite{scoville16}). We consider 12 QGs with direct ALMA estimates. From left to right: comparisons of our fiducial values vs. alternative methods to estimate $M_{\rm dust}$, $f_{\rm dust}$, and $\log(\delta_{\rm DGR})$.}
    \label{appfig:dust_comp}
\end{figure}

In this study, we choose as a fiducial method the physically motivated DL07 dust emission library, and to assess potential systematics in $M_{\rm dust}$ estimates, we compare it with two alternative, commonly used methods. First, we apply a single temperature optically thin modified blackbody (MBB) model implemented within \texttt{CIGALE} code. We adopt a fixed dust temperature of $T_{\rm dust} = 21,\mathrm{K}$, representative of QGs (\citealt{magdis2021interstellar}).
We also test the empirical calibration of \citealt{scoville16}, which estimates $M_{\rm dust}$ directly from ALMA flux density at rest-frame $850\:\mu$m using 

\begin{equation}
M_{\rm dust} = \frac{S_{\nu, \rm obs} D_L^2}{\kappa_{\nu_{\rm rest}} (1+z)} \left( B_{\nu_{\rm rest}}(T_{\rm dust}) \right)^{-1},
\end{equation}

where $B_\nu(T)$ is the Planck function evaluated at temperature $T$, and $\kappa_\nu \equiv \kappa_{\nu_0} (\nu/\nu_0)^{\beta}$ is the dust absorption coefficient. We adopt the standard emissivity of $\kappa_{\nu_0} = 2.64\, \rm m^2/kg$ at $125\:\mu$m (\citealt{dunne2011herschel}), and assume $\beta = 2.0$, keeping $T_{\rm dust} = 21\:\mathrm{K}$ to be consistent with literature. We apply both methods to our ALMA-detected QGs and compare the resulting $M_{\rm dust}$, $f_{\rm dust}$ and $\delta_{\rm DGR}$ to our fiducial DL07-based values. As shown in Fig.~\ref{appfig:dust_comp}, all methods yield values generally consistent within $\sim$0.2 dex. The overall agreement supports the robustness of our estimates, though uncertainties remain due to the assumed $T_{\rm dust}$ and limited knowledge about the AGN-presence in our sources. 

\section{Properties of ALMA observed QGs from this study}
\label{appendixC}

\vspace{-0.5cm}
\begin{figure}[tbh!]
    \centering
    \includegraphics[width=0.52\textwidth,clip]{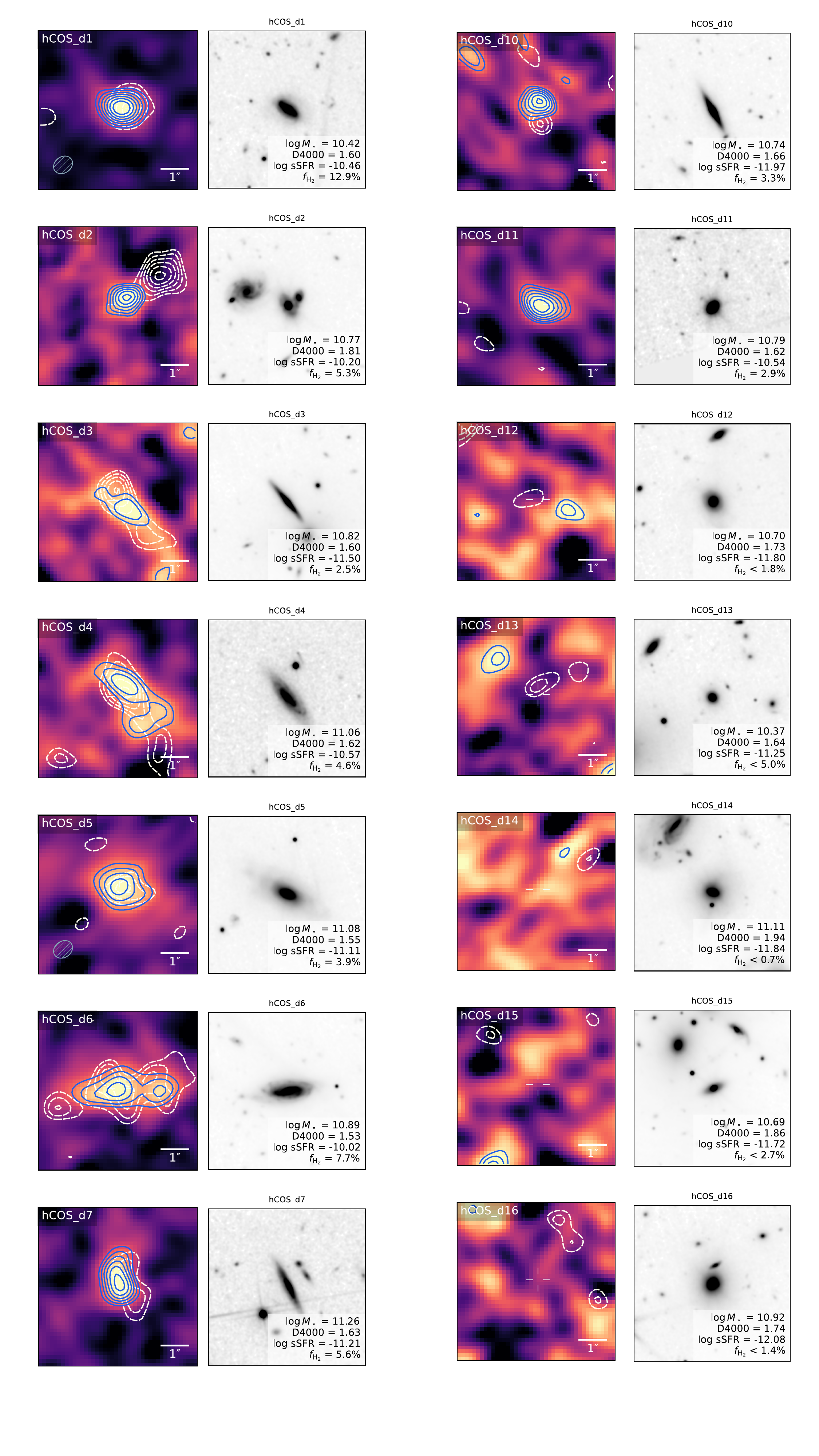}\\
    \vspace{-1.cm}
    \includegraphics[width=0.52\textwidth,clip]{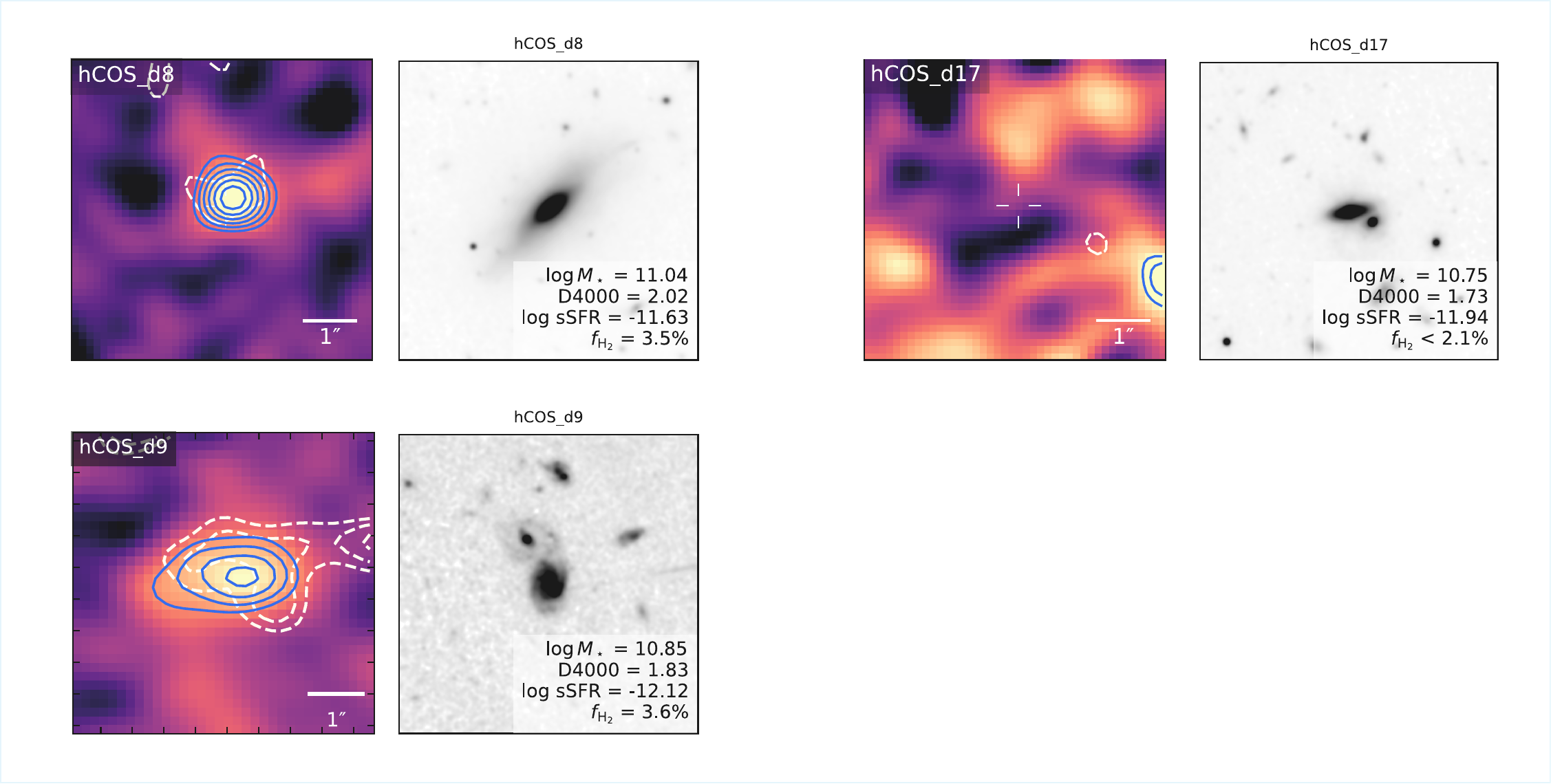}
    \caption{
For each QG: \textit{Left} ALMA Band 6 primary-beam corrected moment-zero map ($5^{\prime\prime}\times 5^{\prime\prime}$ size) overlaid with dust continuum contours (white dashed) and CO(3-2) contours (blue solid) increasing from $2\sigma$ with $1\sigma$ step. Typical beam size of $0.85^{\prime\prime} \times 0.7^{\prime\prime}$ and $1^{\prime\prime}$ scale (corresponding to $\sim5$ kpc) are shown; \textit{Right} Stellar emission from JWST F277W (or HST F160W for sources hCOS-d7 and hCOS-d9) with basic physical properties annotated in the lower-right corner.
}
    \label{appfig:ALMA_cut}
\end{figure}

\begin{table}
\centering
\caption{Summary of physical properties for QGs in this work.}
\resizebox{\textwidth}{!}{
\begin{tabular}{lccclllccccc}
\hline \hline
Name & RA & Dec & $z$ & $\log(M_\star/M_\odot)$ & $\log(\rm Age_{\star}/yr)$ & $\log(\rm sSFR/yr)$ & $S_{\rm 1mm}$ & $S_{\rm CO}$ & $M_{\rm H_2}$ & $M_{\rm dust}$ & $\log(\delta_{\rm DGR)}$ \\
 & [deg] & [deg] &  &  &  &  & [$\mu$Jy] & [Jy km/s] & [$10^9~M_\odot$] & [$10^7~M_\odot$] &  \\
\hline
\vspace{-0.15cm}
\\
hCOS-d1  & 150.350298 & 2.222278 & 0.371 & 10.42 & 9.62 & -10.46 & $66 \pm 11$ & $0.56 \pm 0.02$ & $4.29 \pm 0.15$ & $1.92 \pm 0.29$ & $-2.34$ \\
hCOS-d2  & 150.484972 & 2.082486 & 0.425 & 10.77 & 9.60 & -10.20 & $30 \pm 8$ & $0.39 \pm 0.05$ & $3.84 \pm 0.24$ & $0.89 \pm 0.18$ & $-2.63$ \\
hCOS-d3  & 150.219218 & 2.059187 & 0.371 & 10.82 & 9.70 & -11.50 & $112 \pm 12$ & $0.27 \pm 0.04$ & $1.65 \pm 0.11$ & $3.82 \pm 0.31$ & $-1.64$ \\
hCOS-d4  & 149.956582 & 1.788010 & 0.369 & 11.06 & 9.60 & -10.57 & $97 \pm 15$ & $0.89 \pm 0.04$ & $5.31 \pm 0.14$ & $4.96 \pm 0.39$ & $-2.03$ \\
hCOS-d5  & 149.920863 & 2.031208 & 0.356 & 11.08 & 9.74 & -11.12 & $104 \pm 24$ & $0.85 \pm 0.07$ & $4.73 \pm 0.36$ & $2.58 \pm 0.60$ & $-2.26$ \\
hCOS-d6  & 149.845755 & 1.974381 & 0.338 & 10.89 & 9.58 & -10.02 & $284 \pm 10$ & $1.20 \pm 0.03$ & $6.01 \pm 0.15$ & $6.66 \pm 0.24$ & $-1.96$ \\
hCOS-d7  & 150.602639 & 2.118054 & 0.374 & 11.26 & 9.60 & -11.21 & $54 \pm 11$ & $1.23 \pm 0.06$ & $7.59 \pm 0.22$ & $1.43 \pm 0.29$ & $-2.73$ \\
hCOS-d8  & 150.348257 & 1.948979 & 0.347 & 11.04 & 9.95 & -11.63 & $91 \pm 11$ & $0.72 \pm 0.08$ & $3.80 \pm 0.18$ & $2.90 \pm 0.27$ & $-2.12$ \\
hCOS-d9  & 149.585279 & 2.557561 & 0.388 & 10.85 & 9.90 & -12.12 & $47 \pm 19$ & $0.39 \pm 0.06$ & $2.57 \pm 0.40$ & $1.29 \pm 0.52$ & $-2.30$ \\
hCOS-d10 & 150.386240 & 2.222691 & 0.362 & 10.74 & 9.56 & -11.94 & $93 \pm 8$ & $0.35 \pm 0.03$ & $2.41 \pm 0.18$ & $5.12 \pm 0.21$ & $-1.49$ \\
hCOS-d11 & 149.790964 & 1.919023 & 0.419 & 10.79 & 9.74 & -10.54 & $<21$ & $0.48 \pm 0.04$ & $3.78 \pm 0.32$ & $<0.5$ & $< -2.77$ \\
hCOS-d12 & 150.224869 & 2.236685 & 0.362 & 10.70 & 9.52 & -11.80 & $<42$ & $<0.13$ & $<0.90$ & $<0.50$ & $-$ \\
hCOS-d13 & 149.871032 & 2.233226 & 0.345 & 10.37 & 9.49 & -11.25 & $121 \pm 22$  & $<0.12$ & $<1.18$ & $3.48 \pm 0.30$ & $>-1.53$ \\
hCOS-d14 & 150.182112 & 2.113393 & 0.371 & 11.11 & 9.85 & -11.84 & $<28$ & $<0.13$ & $<0.90$ & ($0.17 \pm 0.12$) & $(> -2.74)$ \\
hCOS-d15 & 150.094894 & 2.297028 & 0.360 & 10.69 & 9.52 & -11.72 & $<84$  & $<0.2$ & $<1.5$ & $<0.3$ & $-$ \\
hCOS-d16 & 150.184392 & 2.067280 & 0.346 & 10.92 & 9.62 & -12.08 & $<84$  & $<0.2$ & $<1.5$ & $<0.3$ & $-$ \\
hCOS-d17 & 149.875440 & 2.200613 & 0.345 & 10.75 & 9.56 & -11.94 & $<84$  & $<0.2$ & $<1.5$ & $<0.3$ & $-$ \\
\hline
\vspace{-0.1cm}
\label{apptable:sourcelist}
\end{tabular}}

\parbox{\textwidth}{
\footnotesize
\textbf{Notes.} RA and Dec correspond to J2000 coordinates; $S_{\rm 1mm}$ is the observed integrated ALMA flux density; $S_{\rm CO}$ is the velocity-integrated line flux; $M_{\rm H_2}$ is derived assuming $\alpha_{\rm CO} = 4.36$; $\log(\delta_{\rm DGR})$ denotes the logarithmic dust-to-gas ratio, defined as $\log(M_{\rm dust}/M_{\rm mol})$; Values in parentheses correspond to source for which $M_{\rm dust}$ was estimated by fitting the DL07 dust model using available archival photometry from de-blended low-resolution \textit{Spitzer} MIPS and \text{Herschel} PACS/SPIRE images, complemented with ALMA upper limits from this study. Given the significant uncertainties associated with this method, these values should be interpreted with caution. 
}
\end{table}

\end{document}